\documentclass[myaps,myrmp,twocolumn]{myrevtex4}
\usepackage{graphicx}% Include figure files

\begin{document}

\title{Inverse correlation between quasiparticle mass and $T_c$ in a cuprate high-$T_c$ superconductor}

\author{C. Putzke$^1$}
\author{L. Malone$^1$}
\author{S. Badoux$^2$}
\author{B. Vignolle$^2$}
\author{D. Vignolles$^2$}
\author{W. Tabis$^{2,3}$}
\author{P. Walmsley$^1$}
\author{M. Bird$^1$}
\author{N.E. Hussey$^4$}
\author{C. Proust$^2$}
\author{A. Carrington$^1$}

\affiliation{
$^1$	H.H. Wills Physics Laboratory, University of Bristol, Tyndall Avenue, Bristol, BS8 1TL, UK.\\
$^2$ 	Laboratoire National des Champs Magn\'{e}tiques Intenses (CNRS-INSA-UJF-UPS), 31400 Toulouse, France.\\
$^3$	AGH University of Science and Technology, Faculty of Physics and \\ Applied Computer Science, Al.\ Mickiewicza 30, 30-059 Krakow, Poland.\\
$^4$	High Field Magnet Laboratory (HFML-EMFL), Radboud University, Toernooiveld 7, 6525 ED Nijmegen, the Netherlands.\\
}

\begin{abstract}
Close to a zero temperature transition between ordered and disordered electronic phases, quantum fluctuations can lead to a strong enhancement of the electron mass and to the emergence of competing phases such as superconductivity. A correlation between the existence of such a quantum phase transition and superconductivity is quite well established in some heavy fermion and iron-based superconductors and there have been suggestions that high temperature superconductivity in the copper oxide materials (cuprates) may also be driven by the same mechanism.  Close to optimal doping, where the superconducting transition temperature $T_c$ is maximum in the cuprates, two different phases are known to compete with superconductivity: a poorly understood pseudogap phase and a charge ordered phase.  Recent experiments have shown a strong increase in quasiparticle mass $m^*$ in the cuprate YBa$_2$Cu$_3$O$_{7-\delta}$  as optimal doping is approached suggesting that quantum fluctuations of the charge ordered phase may be responsible for the high-$T_c$ superconductivity.  We have tested the robustness of this correlation between $m^*$ and $T_c$ by performing quantum oscillation studies on the stoichiometric compound YBa$_2$Cu$_4$O$_8$ under hydrostatic pressure.  In contrast to the results for YBa$_2$Cu$_3$O$_{7-\delta}$, we find that in YBa$_2$Cu$_4$O$_8$ the mass decreases as $T_c$ increases under pressure.  This inverse correlation between $m^*$ and $T_c$ suggests that quantum fluctuations of the charge order enhance $m^*$ but do not enhance $T_c$.
\end{abstract}

\maketitle

\section*{Introduction}
In a variety of systems: organics, heavy fermions, iron-pnictides, and copper-oxides (cuprates) superconductivity is found in close proximity to an antiferromagnetic phase (1-4).  Starting with the non-superconducting parent antiferromagnetic phase, as the material properties are tuned by the application of pressure or chemical doping, the magnetic ordering temperature $T_N$ decreases. The doping/pressure where $T_N$ extrapolates to zero is known as a quantum critical point (QCP).  In many systems the superconducting transition temperature $T_c$ is maximum close the doping/pressure where the QCP is located thus suggesting an intimate connection between the two.  Close to a QCP, quantum fluctuations between the ordered magnetic phase and the disordered paramagnetic phase become strong and if such fluctuations produce electron-pairing this may enhance $T_c$. The strong fluctuations also affect the normal-state properties, causing a strong increase in quasiparticle mass and non-Fermi liquid behavior of the transport properties. Hence close to a QCP the normal-state entropy is increased and this too could provide a boost for $T_c$ (1,3).

The applicability of such a quantum critical scenario to the cuprate superconductors is somewhat debatable, as in these materials $T_c$ is zero close to the QCP of the antiferromagnetic phase.  Instead $T_c$ is maximal close to the end points of two other phases: the pseudogap phase and a charge density wave (CDW) phase. Whether there is a QCP at the end points of either of these phases is unclear, as is their exact position (5, 6). The nature and origin of the long established pseudogap phase, which can be characterized as a depression of the density of states at the Fermi level (7) and a decoherence of quasiparticles on certain parts of the Fermi surface (8), is not well understood.  In contrast, the more recently discovered CDW phases (9-11) seems to be more conventional, although many aspects, such as the microscopic mechanism of their formation and their variation with pressure still need to be clarified.  There are two distinct CDW phases. One phase, which is essentially two-dimensional,  forms independent of magnetic field below $T\sim 150$\,K (10, 11), whereas a second distinct three-dimensional order forms for fields above $\sim$\,15\,T and below $T\sim 60$\,K (9, 12, 13).

A CDW produces an additional periodic potential which reduces the size of the Brillouin zone and folds the Fermi surface into small Fermi pockets.  There is compelling evidence that a CDW induced Fermi surface reconstruction is responsible for the small Fermi surface pockets which are observed in quantum oscillation (QO) measurements (14), so QO give a precise way of characterizing the normal state electronic structure and how this is affected by the CDW.  In particular, from the thermal damping of the QO amplitude the quasiparticle mass $m^*$ can be inferred.  $m^*$ is directly related to the renormalization of the band-structure close to the Fermi level coming from the interaction of the electrons with collective excitations such as phonons or spin/CDW fluctuations.

Recently, the variation of $m^*$ in the cuprate YBa$_2$Cu$_3$O$_{7-\delta}$ (Y123) was measured over a wide range of hole doping ($n_p$) (15, 16).  A dome shaped variation of $1/m^*$ versus $n_p$ was found, with $1/m^*$ extrapolating to zero at two points, $n_{p,c1}\simeq 0.08$, and  $n_{p,c2}\simeq 0.18$.  This is suggestive of QCPs close to  $n_{p,c1}$ and $n_{p,c2}$ and as the latter value is close to the value ($n_p\simeq 0.165$) where $T_c$ is maximal (17) this may indicate that the interactions responsible for the mass enhancement also give rise to the high $T_c$ superconductivity (18).  It is natural to assume that these interactions are the fluctuations of the CDW although additional coupling to the pseudogap fluctuations are also a possible cause (15).

Although chemical doping, for example by changing the oxygen content $\delta$ in Y123, is the most widely used method of tuning the properties of cuprates, pressure is another available tuning parameter, and indeed the highest transition temperatures are achieved under high pressure (19).  Here we report measurements of quantum oscillations under high pressure in the stoichiometric cuprate YBa$_2$Cu$_4$O$_8$ (Y124). This allows us to study how the electronic correlations evolve as $T_c$ is increased with pressure in a single sample with fixed stoichiometry.

Y124 has a similar structure to Y123 but with a double CuO chain layer. Unlike Y123, in Y124 the oxygen stoichiometry is fixed and so cannot be used to change the doping or $T_c$.  However, the application of hydrostatic pressure to Y124 leads to a strong increase in $T_c$ of 5.5\,K/GPa at low pressure (20), increasing $T_c$ from 79\,K at ambient pressure (20) [where $n_p \simeq 0.13(1)$ (21)] to $\sim 105$\,K at $\sim$ 8\,GPa  (22, 23) at which point it undergoes a structural phase transition and becomes non-superconducting (24, 25).  While direct (x-ray or NMR) evidence for the existence of a CDW in Y124 is missing up to now, its existence is likely as the transport properties (26) show similar features to those in Y123 which have been ascribed to CDW formation and most importantly, QO with frequency similar to Y123, which is a key signature of the Fermi surface reconstruction caused by CDW formation, have been observed (21, 27).

\section*{Results}

\begin{figure}
\begin{center}
\includegraphics[width=0.85\linewidth]{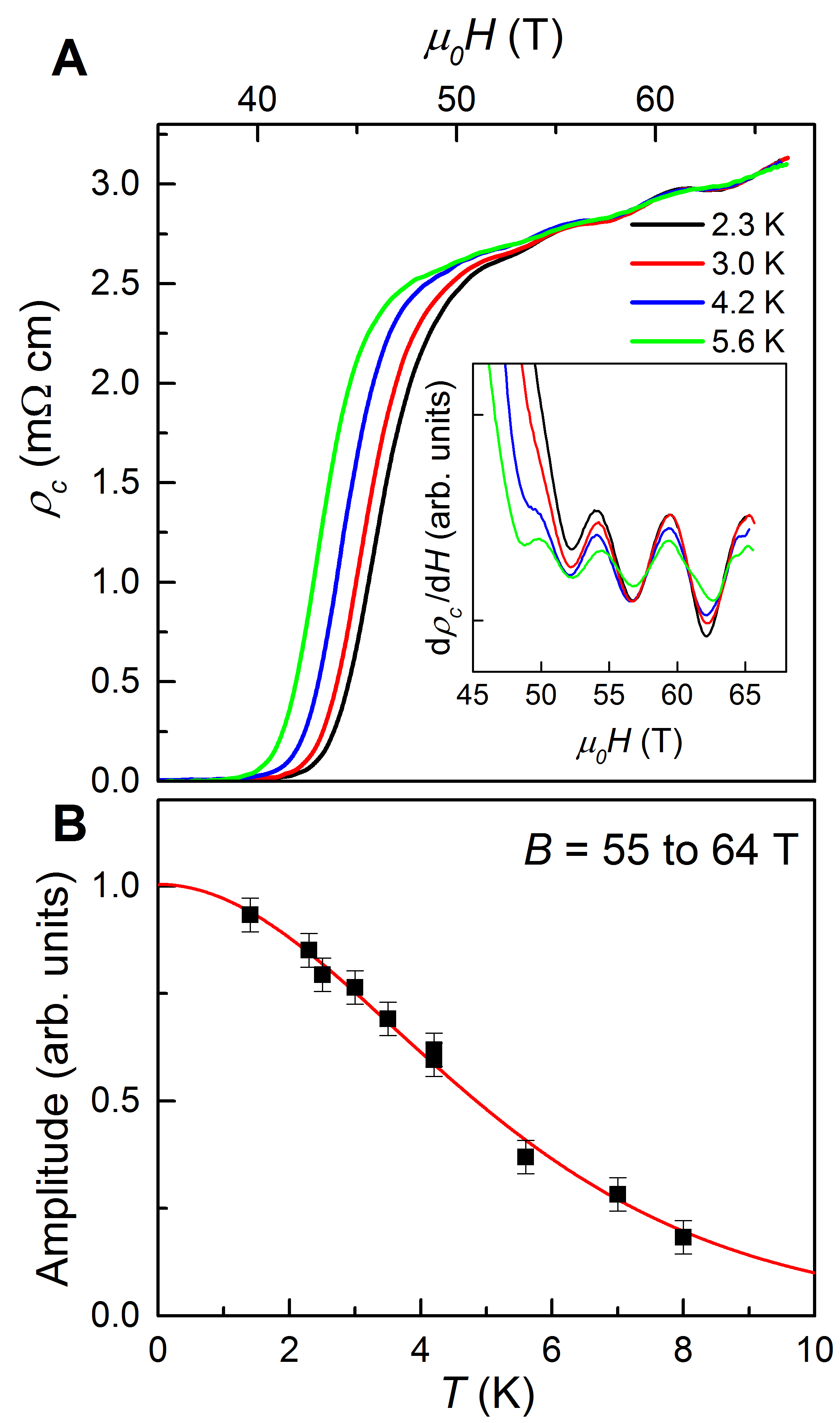}
\caption{Temperature dependent magnetoresistance of Y124. (A) c-axis resistivity at different temperatures, measured up to  $\mu_0H=67$\,T. The inset shows the derivative $d\rho_c/dH$ to emphasize the Shubnikov-de Haas quantum oscillations. (B) The temperature dependence of the quantum oscillation amplitude. The red line shows a fit to the LK expression (see Methods) giving a quasiparticle mass $m^* = 1.80(5) m_e$  (where $m_e$ is the free electron mass).}
\end{center}
\end{figure}

Ambient-pressure measurements of the $c$-axis resistivity $\rho_c$ in pulsed magnetic field up to 67\,T applied parallel to $c$ (perpendicular to the CuO$_2$ planes) of a single crystal sample of Y124 are shown in Fig. 1.  Shubnikov-de Haas quantum oscillations in the resistance are clearly visible in the raw data above $\sim$52\,T, but are more clearly seen in the $d\rho_c/dH$ (inset Fig.1).  The frequency $F$ (in inverse field) of these oscillations is 640\,T consistent with previous reports (21, 27).  The size of the oscillations is remarkably reproducible between different crystals (Fig. S1).  By fitting the temperature dependence of the oscillation amplitude the quasiparticle mass can be extracted (Fig. 1B).  We find a value of $m^*$=1.80(5) (in units of the free electron mass).  A value of $m^*$=1.9(1) was measured in a two further samples (Fig. S2).   This is consistent with another recent study (28) however older measurements (21, 27) found larger values of $m^*$ but the errors were considerably larger than in the present work (see also methods).

\begin{figure}
\begin{center}
\includegraphics[width=1.0\linewidth]{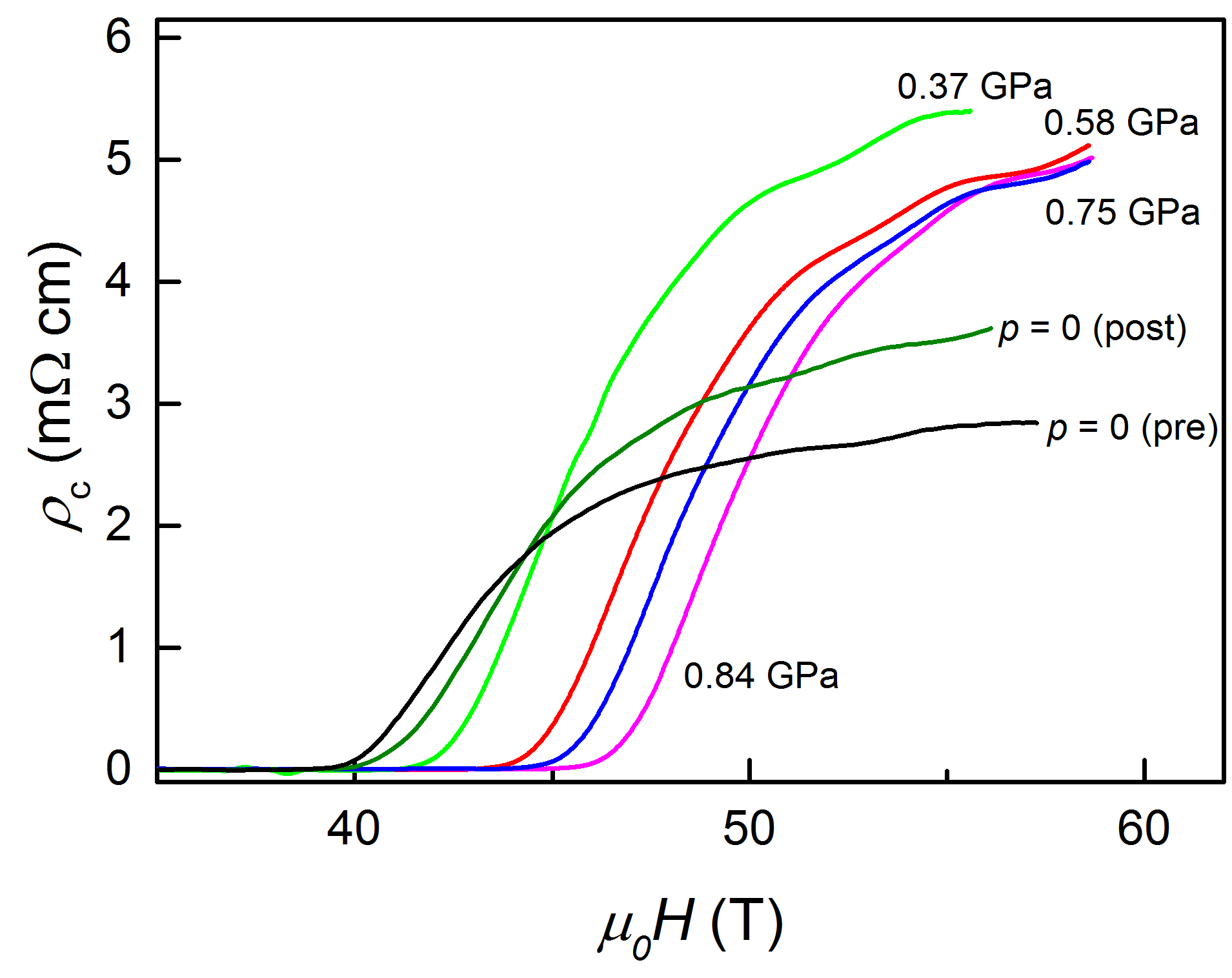}
\caption{Magnetoresistance of Y124 at various pressures at $T$=2.5\,K. Data for $p=0$ are shown both before the pressure was applied and after it was removed.}
\end{center}
\end{figure}

This same sample was then transferred to a non-metallic pressure cell (29). Measurements of $\rho_c(H,T)$ at different applied pressures ($p$) up to $p=0.84$\,GPa are shown in Fig. 2.  Because of the larger bore of the magnet used for the pressure cell measurement, the maximum field here was 58\,T.  At room temperature $\rho_c$ decreases linearly with increasing pressure (Fig. S3A) and this decrease is essentially independent of temperature down to $T_c(H=0)$ (Fig. S3B).  However, when superconductivity is suppressed by field at low temperatures, a dramatically different behavior emerges. As can be seen in Fig. 2, both the magnitude of the resistivity and its slope at 2.5\,K and 50\,T almost doubles between $p = 0$ and 0.37\,GPa (which is the lowest pressure we are able to apply), which suggests that at low temperature, the size of the magnetoresistance has been significantly increased by the application of pressure, presumably due to an increase of the mean-free-path on some part of the Fermi surface. This result was reproduced in 3 samples.

\begin{figure*}
\begin{center}
\includegraphics[width=0.85\linewidth]{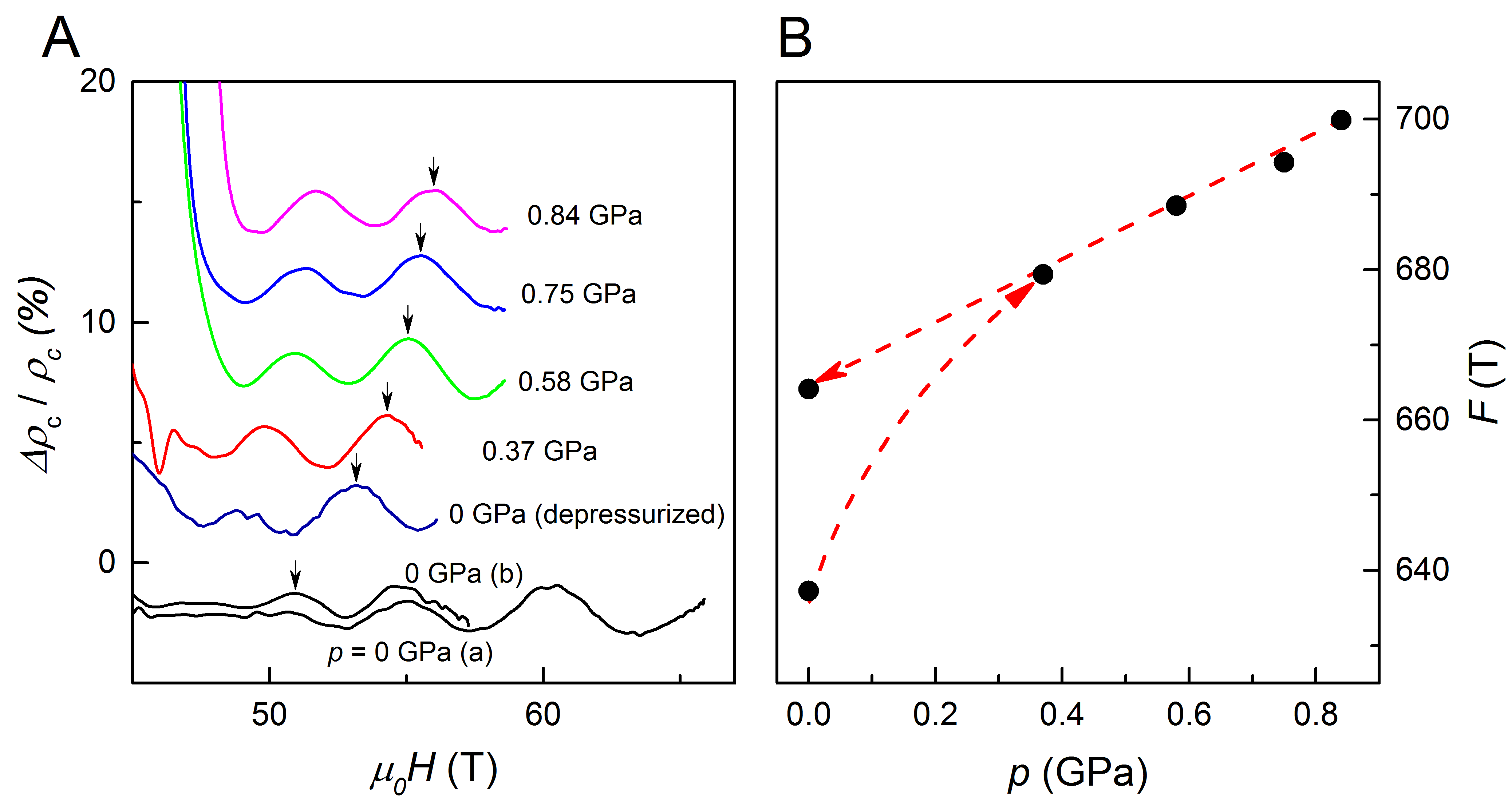}
\caption{Oscillatory part of the resistance versus field for different pressures at $T=2.5$\,K.  (A) For $p=0$\,GPa there are 3 curves, (a) and (b) were measured outside the pressure cell (a) in a 70\,T coil and (b) in the same 60\,T coil as the pressure cell measurements.  The third curve is the result at $p=0$ after depressurizing the cell.  The arrows mark the position of a local maximum $B_{max}$ in $\Delta \rho_c/\rho_c$. The curves have been offset vertically for clarity. (B) Evolution of the QO frequency with pressure. For $p=0$ (before pressurization) the frequency was taken from a direct fit to $\sin(2\pi F/B+ \phi)$, then the changes in the frequency as $p$ is varied are inferred from $B_{max}$.  Similar changes in $F$ were also found by fitting each curve (as for $p=0$) but with a higher noise level.}
\end{center}
\end{figure*}

We were able to observe quantum oscillations at all pressures. The oscillatory part of the magnetoresistance at a fixed temperature of 2.5\,K is shown in Fig. 3A. The relative oscillation amplitude ($\Delta \rho_c/\rho_c$) increases by around a factor 2 between $p=0$ and 0.37\,GPa but then remains approximately constant as pressure is increased further.  There is a shift in the phase of the oscillations as a function of $p$ which is consistent with a small shift of frequency, equivalent to the Fermi surface area changing by $\sim 10\%$ between 0 and 0.84\,GPa (Fig. 3B).  Again the change is largest between 0 and 0.37\,GPa and then varies linearly for further increase in pressure.  On depressurization to $p=0$ after the highest pressure,  $\rho_c(T)$ for $T > T_c (H=0)$ approximately returns to the values before pressurization (Fig. S3) as does $T_c$, but the magnetoresistance at low temperature (Fig. 2) remains larger.  The phase shift in the depressurized ($p$=0) state follows the linear relation found at higher pressures (Fig. 3B).

\begin{figure*}
\begin{center}
\includegraphics[width=0.85\linewidth]{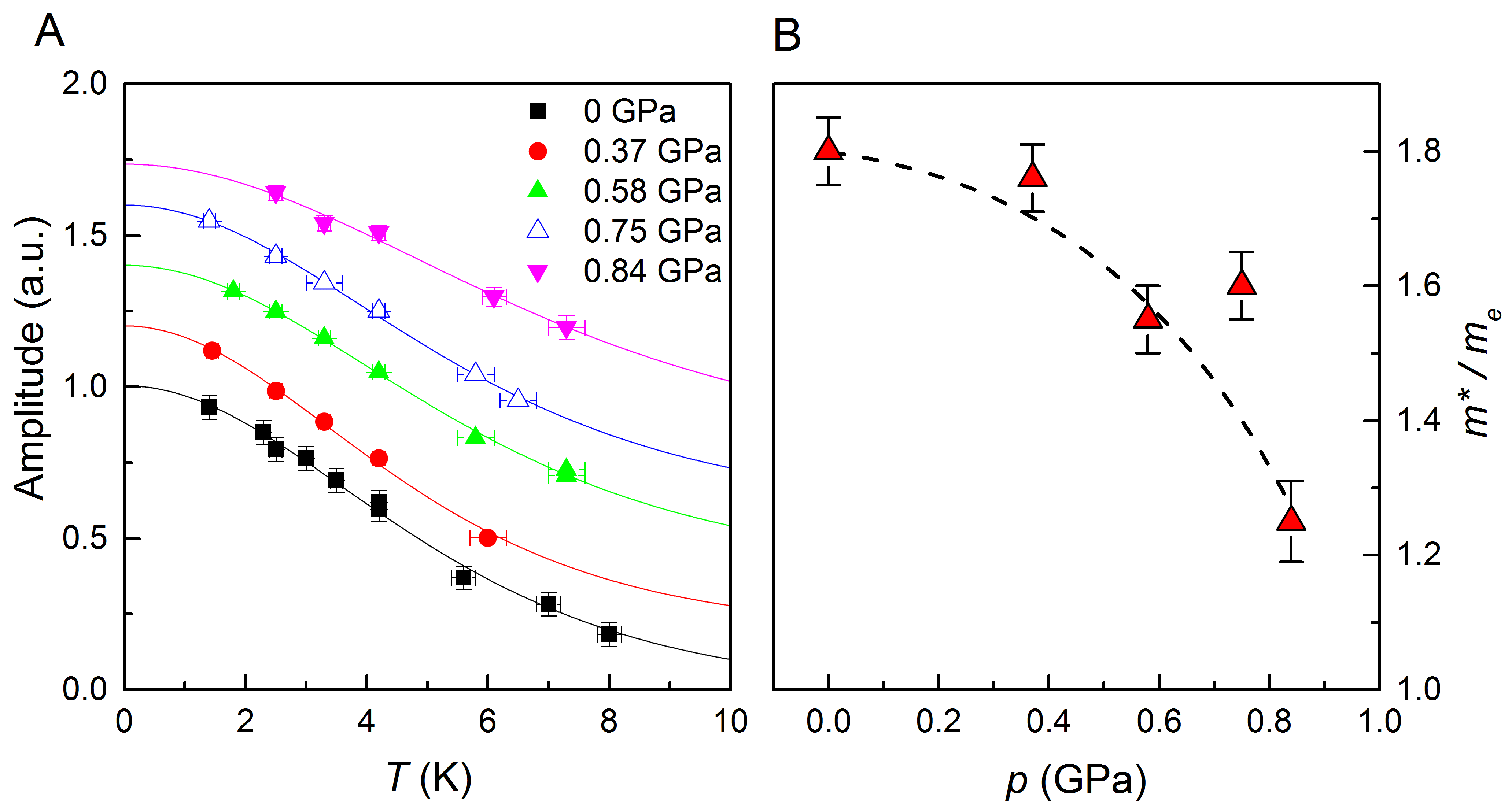}
\caption{The quasi-particle effective mass in Y124 under hydrostatic pressure. (A) Amplitude of quantum oscillations as a function of temperature. The curves have been offset vertically for clarity. The solid lines are fits to the LK formula. The field windows used are given in Table S1.  (B) Variation of the quasiparticle mass with pressure extracted from the fits. The dashed line is a guide to the eye.}
\end{center}
\end{figure*}

The temperature dependence of the oscillation amplitudes at the different pressures, along with the fits to determine the effective masses are shown in Fig. 4A.  The mass found at low pressures measured inside the pressure cell is the same as the value measured outside the cell where the sample is in contact with liquid helium at low temperatures. This shows there are no systematic errors in the mass determination introduced by the pressure cell.  Fig. 4B shows a central result of this paper, namely that the effective mass strongly decreases with pressure and hence has an inverse correlation with $T_c$.

\section*{Discussion}

In Fig. 5, we make a comparison between the effective mass and $T_c$ in Y123 where $T_c$ is varied by changing oxygen content $\delta$ and in Y124 where $T_c$ is varied by pressure.  The inverse of the mass is plotted to emphasize the almost linear relation between ($1/m^*$) and $n_p$ found in Y123 close to the two putative quantum critical points (15).  It is clear that in Y124 the opposite trend is observed, so that the effective mass decreases as $T_c$ is increased towards the maximum.

\begin{figure}
\begin{center}
\includegraphics[width=0.85\linewidth]{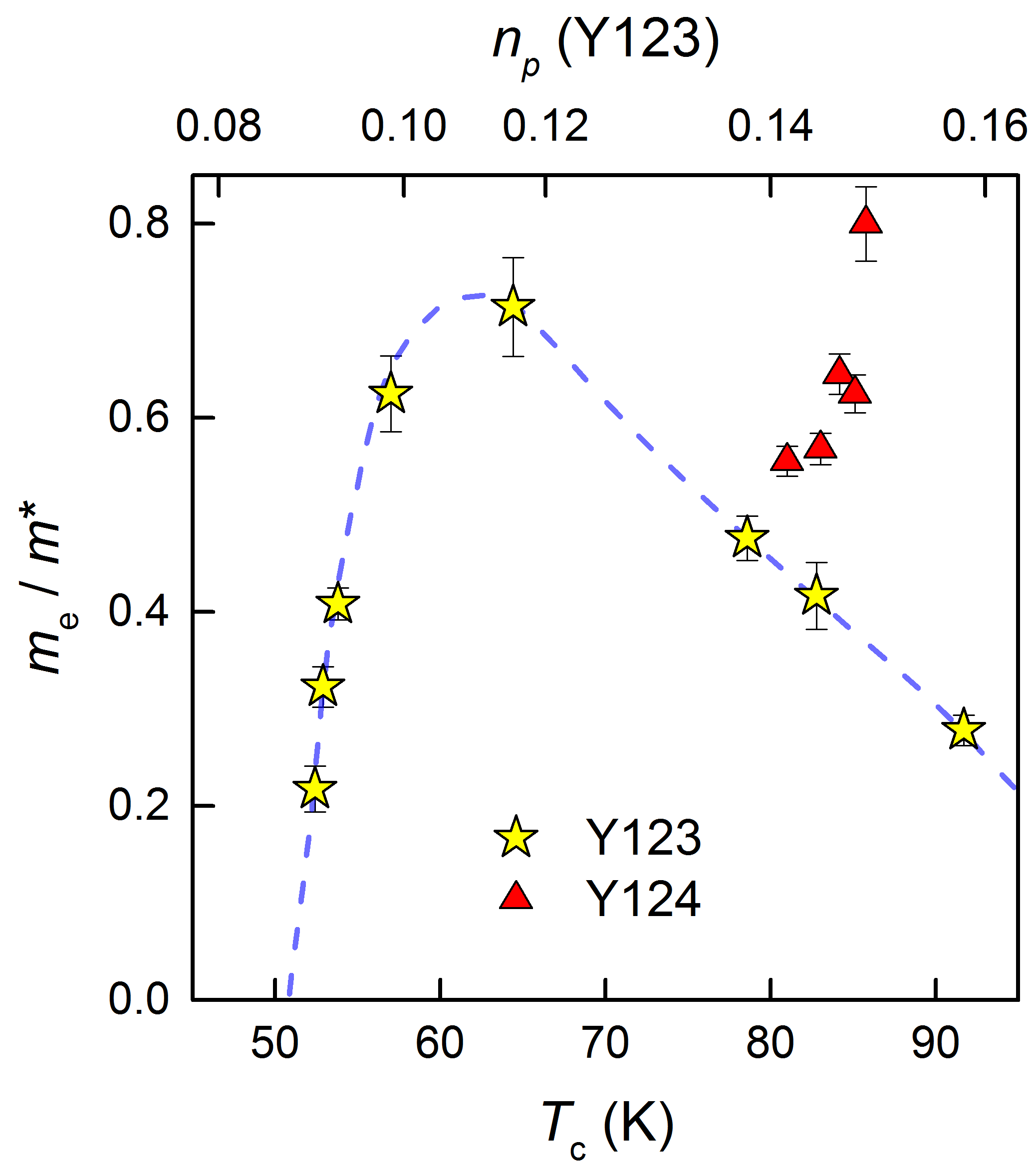}
\caption{Quasiparticle mass of Y124 compared to Y123 plotted versus $T_c$.  The top scale shows the  hole doping level for Y123 (17).  Data for Y123 is taken from Ref. (15) and references therein. The dashed line is a guide to the eye.}
\end{center}
\end{figure}

To interpret these results it is necessary to discuss both how the physical properties of Y124 are changed under pressure and also the mechanism of mass enhancement.  The main effect of hydrostatic pressure on the structure of Y124 is to decrease all three lattice parameters, but the distance from the CuO$_2$ plane to the apical oxygen reduces twice as rapidly as the $c$-axis parameter itself (30).  This distance is correlated with $n_p$ in Y123 (17, 31) and hence suggests that $n_p$ increases with increasing pressure in Y124 (30).   This is supported by experiments which show that with increasing pressure the room temperature resistivity and thermoelectric power (32, 33) both decrease as does the pseudogap temperature as measured by NMR Knight shift  (34).   It is also supported by our observation that the QO frequency in Y124 increases with pressure.  In Y123, an approximately linear increase in $F$ with increasing $n_p$ (35) has been found and it is interesting to note that when plotted against $T_c$ the data for $F(p)$ for Y124 form an almost perfect linear continuation of the increase in $F$ with $n_p$ for Y123 (Fig. S4).

Quantitative estimates of $dn_p/dp$ (see Methods) and the fact that the maximum $T_c$ in Y124 and Y123 under pressure is $\sim 105$\,K, which is considerably higher than the maximum of $T_c\simeq 94$\,K that can be achieved by changing $\delta$ in Y123 alone, suggest that charge transfer is not the only pressure induced effect which increases $T_c$.   It has recently been proposed that an important additional effect could be a weakening of CDW order with pressure (36).  As the CDW competes for states with superconductivity, as evidenced by the observed dip in $T_c$ and upper critical field $H_{c2}$ at the value of $n_p$ where the CDW is strongest (37), then if the CDW is weakened,  $T_c$ may be enhanced.   Although weakening of the CDW order remains a possibility at higher pressures, we find no decrease in the amplitude of the QO with increasing $p$ (Fig. 3A). This strongly suggests that there is no significant decoherence of the high field CDW for $p < 0.84$\, GPa.  In fact, the increase in QO amplitude with increased $p$ indicates an increase in the CDW coherence.

As $T_c$ increases linearly with $p$ ($<$4\,GPa) it is unlikely that the anomalously large effect of the initial pressurization on $F$ (Fig. 3B) and the magnitude of the magnetoresistance (Fig. 2) is due mainly to charge transfer.  More likely it results from an increase in order of any residual vacancies in the material.  This would lead to an increase in the mean-free-path, and in turn, the QO amplitude and the size of the magnetoresistance.  When the pressure is removed this increased order may persist giving rise to the observed hysteresis of $F(p)$ (Fig. 3B) and the magnetoresistance.  We find however, that this increased order does not have any significant effect on $m^*$ (Fig. 4) and so does not affect our central conclusion.

The increase in $m^*$ in Y123 close to optimal doping (15) could be interpreted as resulting from enhanced coupling to fluctuations in either CDW or pseudogap order as these modes would be expected to soften close to their end points. Our key finding is that in Y124 $m^*$ decreases as $T_c$ increases under pressure, which clearly demonstrates that the enhancement found for $m^*$ in Y123 as $\delta$ is reduced, increasing $T_c$ towards optimal, is not universal.   Rather it suggests that the approximate coincidence of $n_{p,c2}$ (the doping where $1/m^*$ extrapolates to zero in Y123) and the doping where $T_c$ is maximum is accidental, so there is no causal connection between the fluctuations giving rise to the mass enhancement and $T_c$.  The decrease in $m^*$ in Y124 could be interpreted as resulting from a decrease in CDW fluctuations caused by the pressure-induced structural changes.  For example, the reduction in $c$-axis length caused by increasing pressure might help stabilize the high field CDW phase where three dimensional coupling is important and hence increase the CDW coherence, thereby reducing the mass enhancing fluctuations.  Elucidation of the microscopic changes will require a structural study of the CDW at high pressure.   Our results suggest that the proximity of the CDW end point to the maximum in $T_c$ with doping is coincidental and that therefore quantum fluctuations of the CDW order do not boost $T_c$ in the cuprates.

\section*{Materials and Methods}
Crystals Growth and Characterization. Crystals of Y124 were grown from mixture of Y123 powder and CuO$_2$ in a KOH-flux (38). The obtained crystals had a typical size of 200$\times$200$\times$100$\mu m^3$ with the shortest dimension being along the $c$-axis.  The composition was checked using x-ray diffraction and lattice parameters were found to be in agreement with previous reports.  The samples displayed a sharp superconducting transition with $T_c$=79\,K.  For resistance measurements, four contacts were made, two each on the top and bottom faces, using evaporated gold pads which were annealed onto the sample at 500$^\circ$C and then contacted with Dupont 4929 silver epoxy.  Typical contact resistances were $<0.1\Omega$   at 4.2\,K.  In zero field, resistance was measured using a standard low frequency lock-in-amplifier method, and in pulsed field using a numerical lock-in digitization technique operating at typically 54\,kHz.  Resistance versus temperature curves are shown in Fig. S3B.  $R_c$ shows metallic behavior at low temperature and is similar to the results reported in Ref. (39).

\textbf{Pressure cell}.   A non-magnetic pressure cell was used in the pulsed magnetic field system (29). Because of the limitation of space inside this cell, pressure was determined using the samples themselves as a manometer.  Changes in $R_c$ of the Y124 samples were measured in the non-metallic cell at room temperature and these results were compared to measurements done in a different larger cell where the pressure was measured at low temperature using a Pb manometer and at room temperature using the resistance of a manganin wire (40). In this second cell it was verified, by $c$-axis resistivity measurements, that d$T_c/dp$ was 5.5(5)\,K/GPa in agreement with previous measurements (20).

\textbf{Effective mass determination}: Quantum oscillation data was analyzed by fitting the following function to the $R_c(B)$ data

\[
R(B) = A B^{1/2}\exp(-\alpha/B) \sin (2\pi F/B + \phi ) +P(4)
\]
where $P(4)= a_0 + a_1B + a_2B^2 + a_3B^3 + a_4B^4$.  The first part of the function represents the field dependent part of the Lifshitz-Kosevitch formula (41) for the quantum oscillation amplitude in a three dimensional metal at $T=0$. We fit this to a short section of data (field ranges for each pressure are given in Table S1) at the lowest temperature (typically $T=1.4$\,K) and determine $\alpha$, $F$ and $\phi$.  These parameters are then left fixed and the oscillation amplitude $A$ was then determined as a function of temperature.  The fourth order polynomial $P(4)$ was used to subtract the non-oscillatory background resistance.  The temperature dependence of $A$ was then fitted to the thermal damping factor $R_T=X/\sinh(X)$, $X=2\pi^2k_Bm^*T/(e\hbar B)$ to extract the effective mass $m^*$.  In this formula $B$ is set to the average inverse field in the field range fitted.

\textbf{Thermometry in pulsed magnetic field}:  In order to accurately determine the mass it is important to be able to accurately measure the sample temperature.  This is not trivial in a pulsed field system because the relatively high sample currents needed to get sufficient signal to noise ratio produce heating at the sample contacts.  These effects were minimized in our measurements by obtaining very low contact resistances (around $0.1\Omega$) and having the sample in direct contact with liquid helium (for $T < 4.2$\,K) for the measurements outside of the pressure cell.
For measurement inside the pressure cell the thermometer was mounted directly outside the cell.  Tests using a miniature GeAu thermometer mounted inside the cell, attached to the sample showed that the two thermometers were in thermal equilibrium provided that the resistive heating at the sample contacts was sufficiently low.  The maximum current used for the resistivity measurements was selected so that this was always the case.
The strong increase in $H$ of the superconducting-normal transition with temperature was used as a second check of the temperature. The irreversibility field, $H_{irr}$, defined as the field where the resistance first increases above the noise, varies linearly with $T$ (Fig. S5).  The continuity of this linear behavior when the sample is inside the cell provides an important check that the sample temperature has been correctly determined.

\textbf{Estimates of change of doping with pressure}: A quantitative estimate of the change in $n_p$ with $p$ can be made using the magnitude of the room temperature $a$-axis thermoelectric power ($S_{290}^a$). Using the data for Y124 in Ref.\ (32) along with the universal relation between $n_p$ and $S_{290}^a$ (33, 42) we obtain $dn_p/dp =2.3(6)\times 10^{-3}$ holes/GPa. A second estimate can be made using the pressure dependent NMR Knight shift $^{17}$O data for Y124 from Ref.\ (34).  Following Ref.\ (43) we scale the temperature axis by the pseudogap energy $E_g$ so that the data for different pressures collapses onto a single curve and $^{17}$O($T/E_g$) is approximately constant for $T/E_g > 1$.  For $p$=0 we estimate $E_g=350$\,K which is consistent with the $a$-axis resistivity $\rho_a(T)$ data (39).  From this we obtain $dn_p/dp = 3.6(4)\times 10^{-3}$ holes/GPa, for $p < 4.2$\, GPa which is in reasonable agreement with the above estimate from $S_{290}^a$.   Taking an average of these two values and combining them with an estimate of d$T_c/dn_p$ taken from Y123 at $n_p=0.13$ (17), would predict that for Y124 d$T_c/dp$=2.7(5) K/GPa, i.e. around 2 times smaller than the measured d$T_c/dp$=5.5(1) K/GPa (23).

Although the quantitative precision of this estimate should be viewed with caution because the universal relations between $S_{290}^a$ and $n_p$ (33, 42) and $E_g$ and $n_p$ (43) may not hold under finite pressure it is likely there are other effects which increase $T_c$ in addition to charge transfer.

In simple systems, the QO frequency $F$, which is proportional to the Fermi surface extremal cross-section, can often be used to precisely determine the doping level. In cuprates however, this has so far only been successfully demonstrated in the overdoped cuprate Tl$_2$Ba$_2$CuO$_{6+\delta}$    (44-46).  For the underdoped materials the reconstruction of the Fermi surface complicates the analysis as the area of the reconstructed pocket will depend on the unknown details of the reconstructing potential as well as $n_p$.  In Y123, an approximately linear increase in $F$ with increasing $n_p$ (35) has been found. If we assume that $dF/dn_p$ in Y124 is the same as in Y123 for $0.09<n_p<0.125$ then the slope $dF/dp$ from the linear part of Fig. 3B would give $dn_p/dp =1.0(2)\times 10^{-2}$ holes/GPa which is around 3 times larger than the estimate from $S_{290}^a$. However, as it is not known how universal $F(n_p)$ is in the cuprates this estimate is probably the most uncertain.

\section*{Acknowledgments}
We thank John Cooper, Sven Friedemann, Jeff Tallon, and Louis Taillefer for useful discussions. This work was supported by the Engineering and Physical Sciences Research Council (Grant No. EP/K016709/1).  A portion of this work was performed at the LNCMI, which is supported by the French ANR SUPERFIELD, the EMFL, and the LABEX NEXT. We also acknowledge the support of the HFML-RU/FOM.

\section*{Data availability} Raw resistance versus field or resistance versus temperature data for all figures can be found at DOI:10.5523/bris.p6wv6g0eoh3t1fd2d4fjujuno.

\newpage
\def\figurename{\sffamily\selectfont Fig. S}
\def\tablename{\sffamily\selectfont Table S}
\setcounter{figure}{0}
\onecolumngrid
\noindent
\newpage
{\sffamily\bfseries\selectfont \Large Supplementary Materials}
\vskip 12pt

%\section*{Supplementary Materials}

\begin{figure*}[h]
\begin{center}
\includegraphics[width=0.7\linewidth]{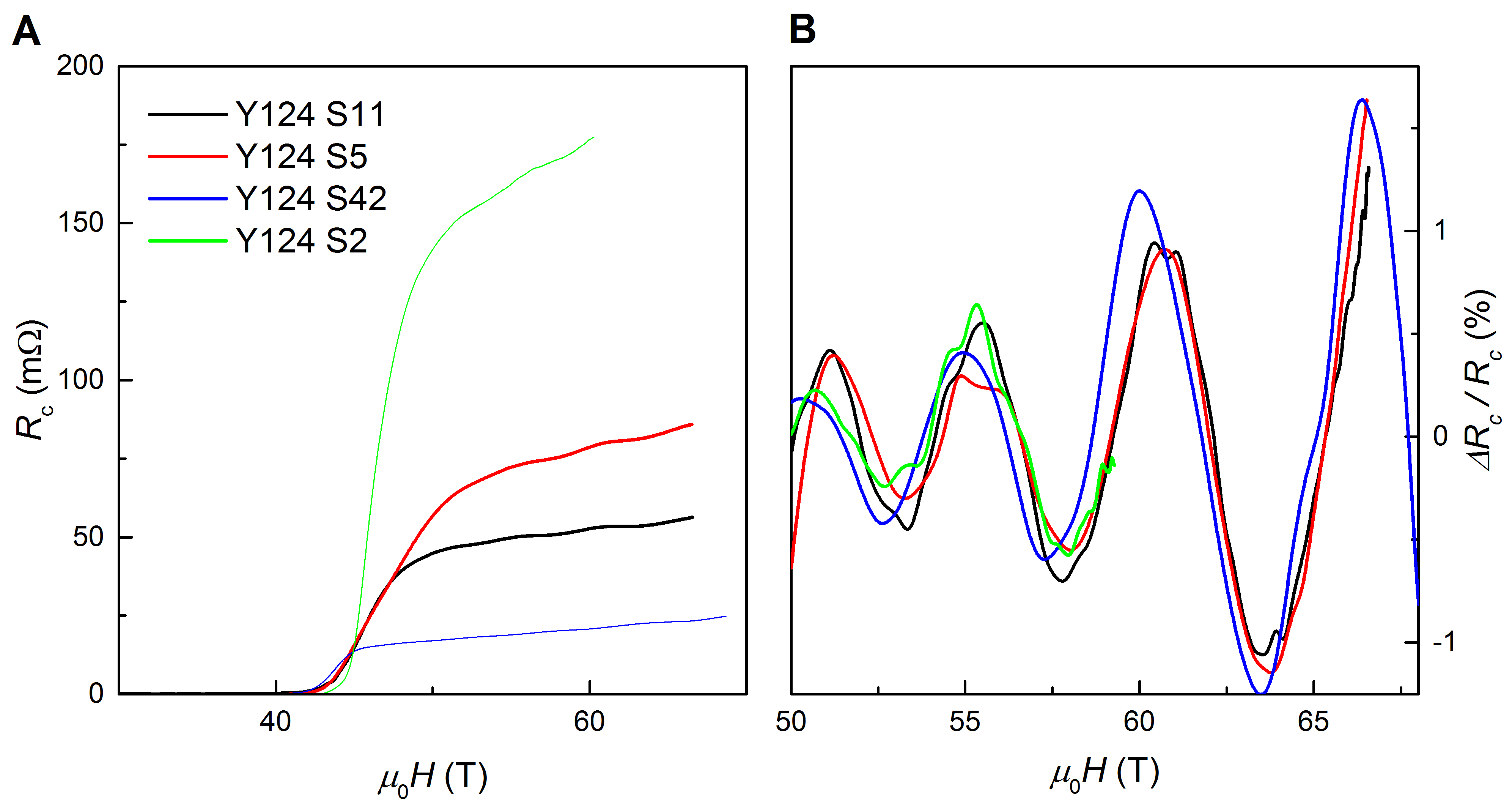}
\caption{Sample dependence of zero pressure quantum oscillations.  (A) $c$-axis resistance versus magnetic field for 4 different samples at $T = 2.5$\,K. (B) The oscillatory part of the resistance after subtracting a smooth polynomial background. The large difference in absolute resistances in A are mainly due to the different dimensions of the samples.}
\end{center}
\end{figure*}

\begin{figure*}[h]
\begin{center}
\includegraphics[width=0.7\linewidth]{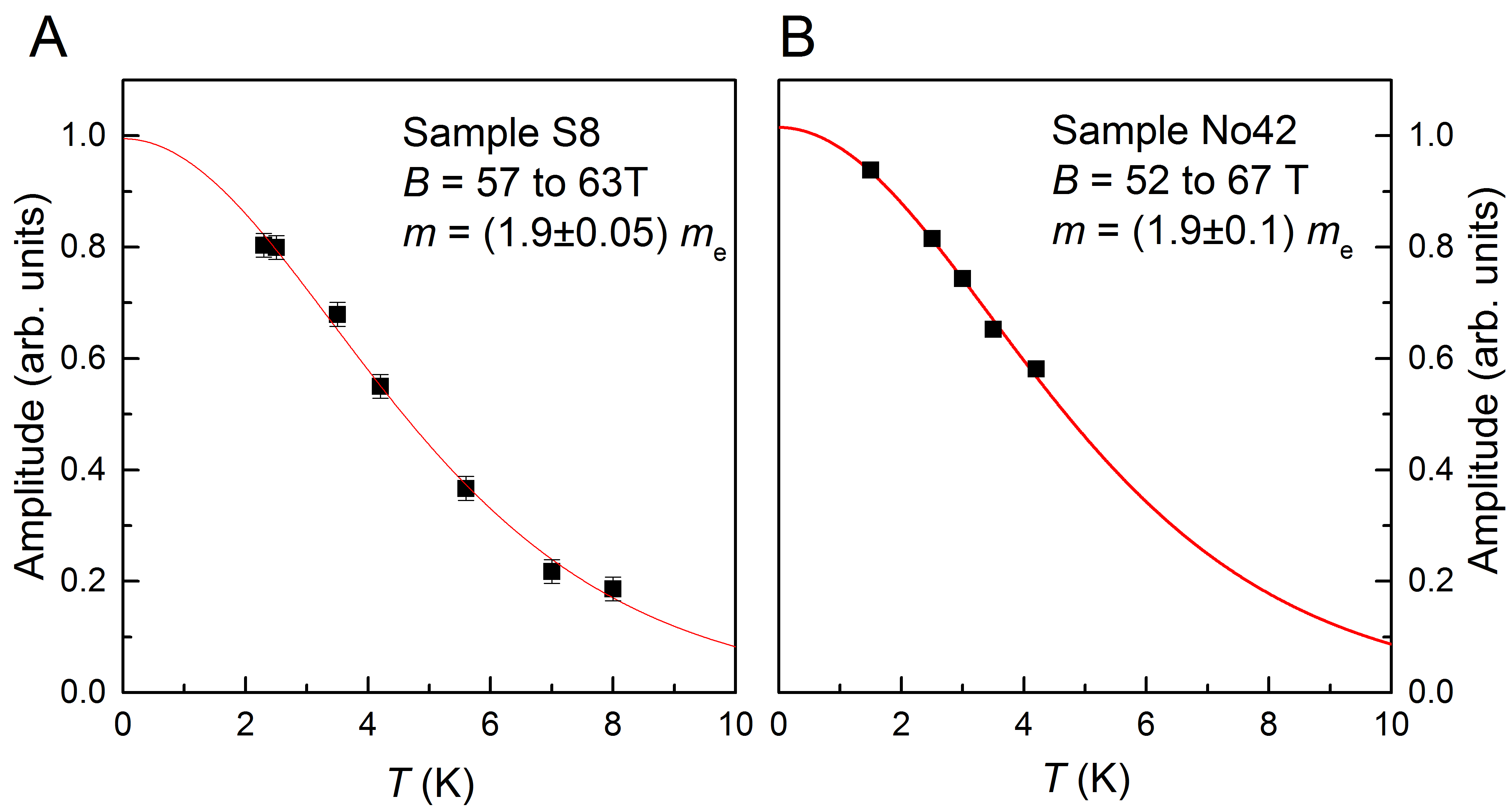}
\caption{Effective mass determination at $p=0$ for two further samples.}
\end{center}
\end{figure*}

\begin{figure*}[h]
\begin{center}
\includegraphics[width=1.0\linewidth]{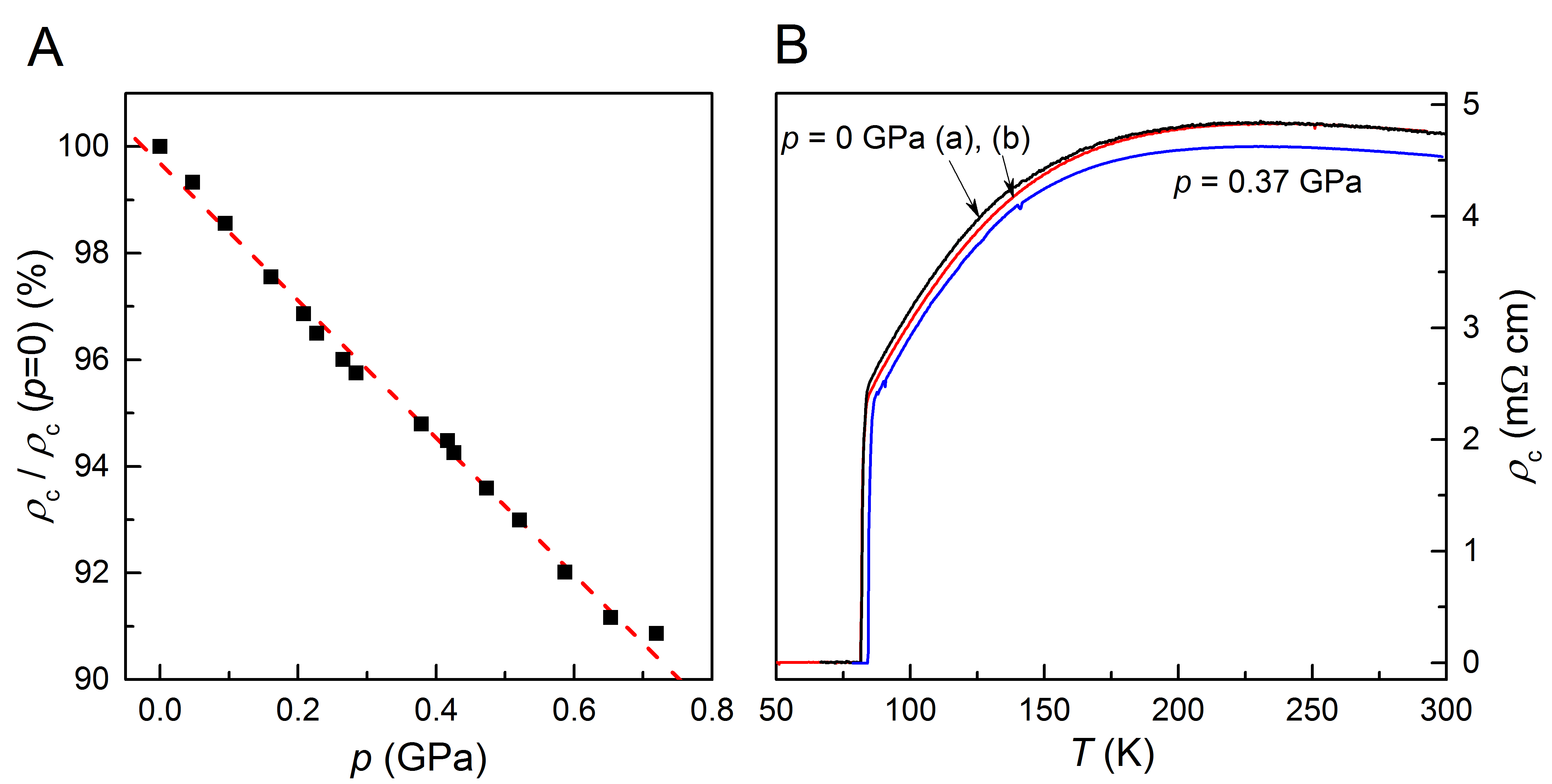}
\caption{Pressure dependence of  $\rho_c(T)$. (A) Relative change in the room temperature $c$-axis resistivity ($\rho_c(T)$) of Y124 sample S11 with pressure. (B) $\rho_c(T)$ at zero pressure (a) before pressurization and (b) after depressurization and also at $p=0.37$\,GPa.}
\end{center}
\end{figure*}

\begin{figure*}[h]
\begin{center}
\includegraphics[width=0.8\linewidth]{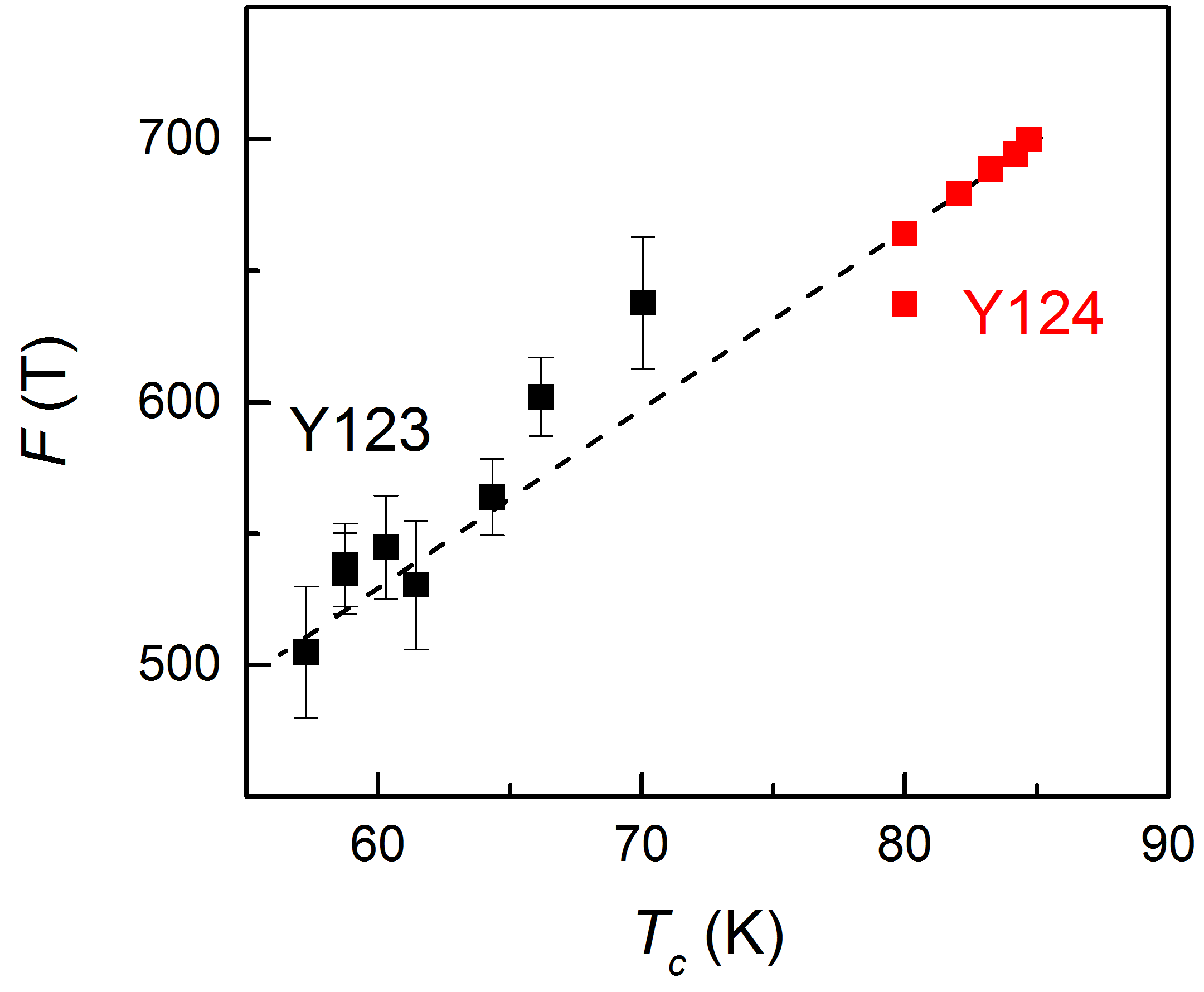}
\caption{Comparison of the changes in quantum oscillation frequency in Y123 and Y124:  Data for Y124 are the same as shown in Figure 3.  For Y123 where the oxygen content $\delta$  has been varied to change the doping state and hence $T_c$, the data were taken from Ref. (35).}
\end{center}
\end{figure*}

\begin{figure*}[h]
\begin{center}
\includegraphics[width=0.6\linewidth]{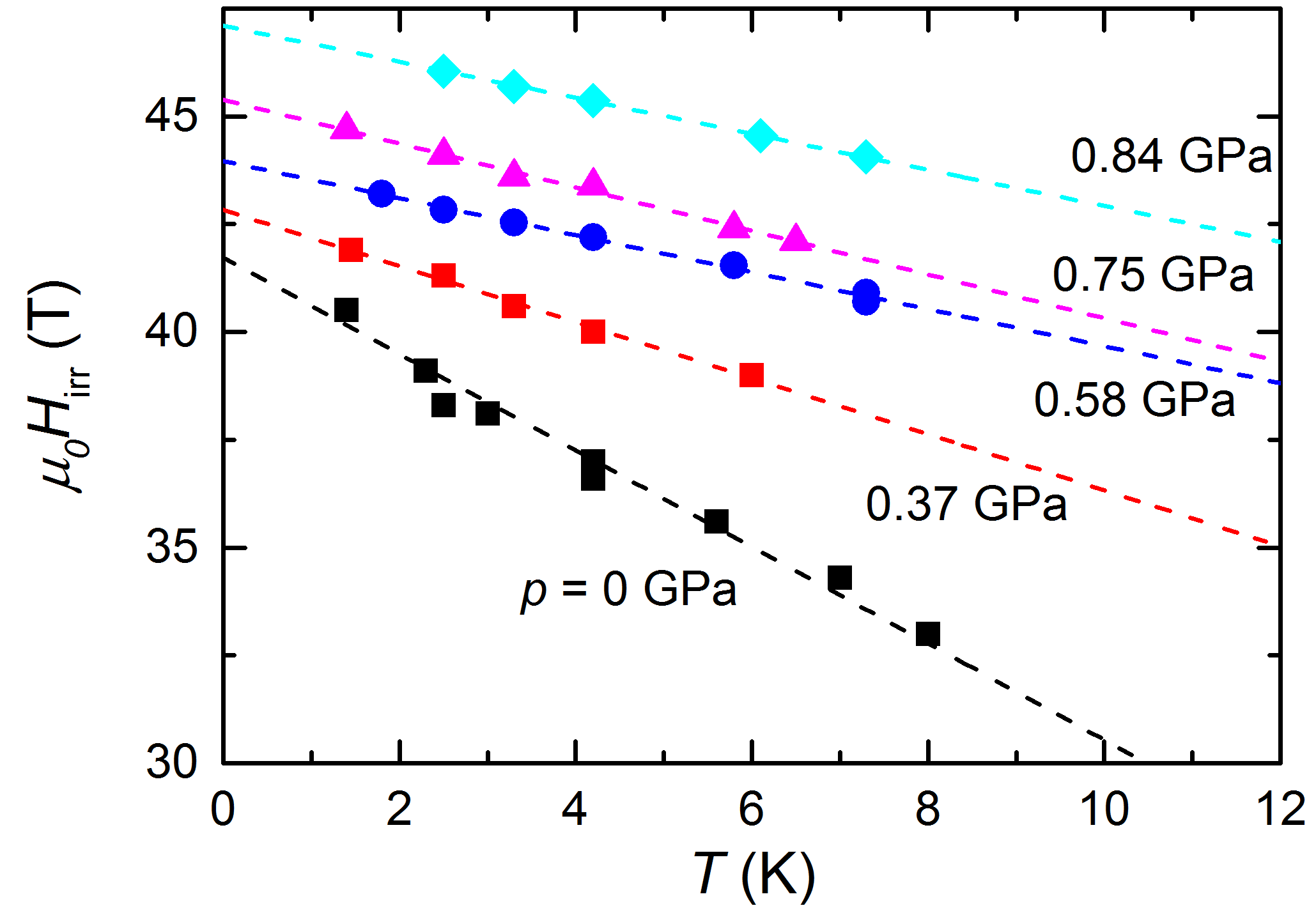}
\caption{Temperature and pressure dependence of the irreversible field $H_{irr}$  The data at $p=0$ were taken outside the pressure cell.}
\end{center}
\end{figure*}

\begin{table*}[h]
\caption{Field windows used for the fitting of the temperature dependent quantum oscillation amplitudes at various pressures as shown in Fig 4.}
\begin{tabular}{|c|c|c|}
\hline
Pressure (GPa)&Field min (T)&Field Max (T)\\
\hline
0	& 55.0&	64.0\\
0.37	&52.0&	55.2\\
0.58	&53.4&	58.7\\
0.75	&53.4&	58.5\\
0.84	&52.5&	58.0\\
\hline
\end{tabular}
\end{table*}
\end{document}